\documentclass[twocolumn,twoside,slac_two]{revtex4}
\usepackage{graphicx}
\usepackage{fancyhdr}
\begin{document}

\title{Dark Matter and Dark Energy from a single scalar field}

%

\author{Roberto Mainini, Loris Colombo, Silvio A. Bonometto}
\affiliation{Dipartimento di Fisica G.~Occhialini, Universit\`a di
Milano--Bicocca, Piazza della Scienza 3, 20126 Milano, Italy
\& I.N.F.N., Sezione di Milano}

\begin{abstract}
The strong CP problem was solved by Peccei \& Quinn by introducing
axions, which are a viable candidate for DM. Here the PQ approach is
modified so to yield also Dark Energy (DE), which arises in fair
proportions, without tuning any extra parameter. DM and DE arise 
from a single scalar field and, in the present ecpoch, are weakly 
coupled. Fluctuations have a fair evolution. The model is also
fitted to WMAP release, using a MCMC technique, and performs
as well as LCDM, coupled or uncoupled Dynamical DE.  Best--fit
cosmological parameters for different models are mostly within
2--$\sigma$ level. The main peculiarity of the model is to favor 
high values of the Hubble parameter.
\end{abstract}

\maketitle

\thispagestyle{fancy}

\section{Introduction}
Axions are likely to be the Dark Matter (DM) that cosmological data
require. Axions arise from the solution of the strong $CP$ problem
proposed by Peccei \& Quinn in 1977 (\cite{1}, PQ hereafter), who
suggested that $\theta$ parameter, in the effective lagrangian term
\begin{equation}
{\cal L}_\theta = {\alpha_s \over 2\pi} \theta \, G \cdot {\tilde G}
\label{eq:n1}
\end{equation}
($\alpha_s$: strong coupling constant, $G$: gluon field tensor),
causing $CP$~violations in strong interactions, is a dynamical
variable. Under suitable conditions $\theta$ approaches zero in our
epoch, so that the term (\ref{eq:n1}) is suppressed, while residual
$\theta$ oscillations yield DM \cite{2,3}.

In the PQ scheme, $\theta$ is the phase of a complex field $\Phi = \phi
e^{i \theta}/\sqrt{2}$; its evolution is set by the potential
\begin{equation}
V(|\Phi|) = \lambda\, [\,|\Phi|^2 - F_{PQ}^2 \, ]^2 ~,
\label{eq:n2}
\end{equation}
whose $U(1)$ invariance breaks when $T<F_{PQ}$ (the PQ energy scale,
which shall be $\sim 10^{12}$GeV). Then $\phi$ settles at the
potential minimum, while $\theta$ takes a random value, different in
different horizons. When chiral symmetry also breaks down, during the
quark--hadron transition, a further term
\begin{equation}
V_1 = \big[\sum_q \langle 0(T)| {\bar q} q |0(T) \rangle m_q \big] 
~(1 - \cos \theta) 
\label{eq:n3}
\end{equation}
must be added to the lagrangian, because of instanton effects. At $T
\simeq 0$, the square bracket approaches $m_\pi^2 f_\pi^2$ ($m_\pi$,
$f_\pi$: $\pi$--meson mass, decay constant).  

The dependence on the site of initial $\theta$'s will causes later
adjustments, as soon as the potential (\ref{eq:n3}) switches on.  When
$\theta$ is small, $V_1 \propto \theta^2$ and the $\theta $ field,
undergoing harmonic oscillations, is DM.

\vglue 0.05truecm 

Here we wish to discuss recent work [4], where the NG potential
(\ref{eq:n2}) is replaced by a tracker potential \cite{5}. Then,
instead of settling on a value $F_{PQ}$, $\phi$ continues to evolve
over cosmological times, at any $T$. As in the PQ case, the potential
involves a complex field $\Phi$ and is $U(1)$ invariant. Its phase is
horizon--dependent but, just as in the PQ case, evolves only after
chiral symmetry breaks, yielding a term similar to (\ref{eq:n3}).

At variance from the PQ case, however, the $\theta$ evolution starts
and continues while also $\phi$ is still evolving. This goes on until
our epoch, when $\phi$ is expected to account for Dark Energy (DE),
while, superimposed to such slow evolution, faster {\it transversal}
$\theta$ oscillations occur, accounting for DM.  This scheme, where
the $\Phi$ fields bears a dual role, with its modulus and phase, will
be dubbed here {\it dual} axion model.

The question is whether $\phi$ and $\theta$ dynamics interfere, still
allowing $\theta$ to take values so small to solve the $CP$ problem
and preserving harmonic oscillations allowing the prescribed amount of
DM. Here we show that the reply is positive. As it can be expected,
however, DM and DE are dynamically coupled, although this coupling
weakens as we approach the present era.

Here we also consider a generalization of the SUGRA potential \cite{5}
as an example of tracker potential. With an energy scale $\Lambda \sim
10^{10}$GeV, this model allows suitable values for today's DM and DE
densities, while $\theta$ is even smaller than for PQ.  DM density
fluctuations are also able to grow and to account for the observed
large scale structure. We fit the expected angular CBR spectra to
data, finding that this model substantially performs as well as
$\Lambda$CDM.

\section{Lagrangian theory}
In the dual-axion model we start from
the lagrangian ${\cal L} =  \sqrt{-g} \{ g_{\mu\nu} 
\partial_\mu \Phi \partial_\nu \Phi   - V(|\Phi|) \} $
which can be rewritten in terms of $\phi$ and $\theta$, adding
also the term breaking the $U(1)$ symmetry, as follows:

\vskip .3truecm
$
{\cal L} = \sqrt{-g} \big\{ (1/ 2)\, g_{\mu\nu} 
\big[\, \partial_\mu \phi \partial_\nu \phi 
+ \phi^2  \partial_\mu \theta \partial_\nu \theta\, \big] -
$

$
~~~~~~~~~~~~
- V(\phi) -m^2(T,\phi) \phi^2 (1 - \cos \theta) 
\big\}  ~.
\label{eq:m1}
$
\vskip .3truecm

\noindent
Here $g_{\mu\nu}$ is the metric tensor. We shall assume that
$ds^2 = g_{\mu\nu} dx^\mu dx^\nu = 
a^2 (d\tau^2 - \eta_{ij}dx_idx_j)$, so that $a$ is the
scale factor, $\tau$ is the conformal time; greek (latin) indeces
run from 0 to 3 (1 to 3); dots indicate differentiation in respect to
$\tau$.  The mass behavior for $T \sim \Lambda_{QCD}$ will
be detailed in Section 3. The equations of motion read
\begin{equation}
\ddot \theta + 2(\dot a/ a+ \dot \phi/ \phi ) ~
\dot \theta + m^2 a^2 \sin \theta = 0~,
\label{eq:m3}
\end{equation}
\vskip -.8truecm
\begin{equation}
\ddot \phi + 2 (\dot a / a)\,  \dot \phi + 
a^2  V'(\phi) = \phi\, \dot \theta^2 ,
\label{eq:m4}
\end{equation}
while the energy densities
$\rho_{\theta,\phi} = \rho_{\theta,\phi;kin} + \rho_{\theta, \phi; pot}$
 and pressures $p_{\theta,\phi} = \rho_{\theta,\phi;kin} -
\rho_{\theta, \phi;pot}$,  under the condition $\theta \ll 1$, are
obtainable from
\begin{eqnarray}	
\rho_{\theta,kin} = {\phi^2 \dot \theta^2 / 2 a^2} ~,~~
\rho_{\theta,pot} =  m^2(T,\phi)  \phi^2 \theta^2/2~,~~
\nonumber
\\
\rho_{\phi,kin} = {\dot \phi^2 /2 a^2} ~,~~
\rho_{\phi,pot} = V(\phi)~.~~~~~~~~~~~
\label{eq:kp}
\end{eqnarray}	

\section{Axion mass}
\begin{figure}[t]
\includegraphics[height=8.truecm,angle=0.]{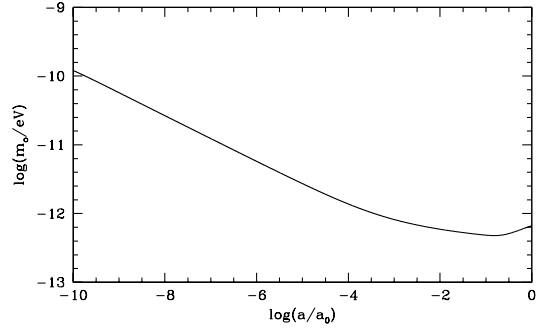}
\vskip-3.2truecm
\caption{$\phi$ variations cause a dependence of the effective axion
mass on scale factor $a$, which is shown here.}
\label{fig:mass}
\end{figure}

According to eq.~(\ref{eq:m3}), the axion field begins to oscillate when
$m(T,\phi)a \simeq 2 (\, {\dot a / a}+{\dot \phi / \phi} \, )$.
In the dual--axion model, just as for PQ, axions become massive when
the chiral symmetry is broken by the formation of the $\bar q q$
condensate at $T \sim \Lambda_{QCD}$. Around such $T$, therefore, the
axion mass grows rapidly. In the dual--axion model, however, a slower
growth takes place also later, because of the evolution of $\phi$.
Then $m(T,\phi)$ is
\begin{equation}
m_o(\phi) = {m_\pi f_\pi/ \phi} =(0.0062 / \phi)~{\rm GeV} ~.
\label{eq:o2}
\end{equation}
Since $\phi \sim m_p$ today, the axion mass is now $m_o \sim 5 \cdot
10^{-13}$eV, while, according to \cite{6}, at high $T:$
\begin{equation}	
m(T,\phi) \simeq 0.1\, m_o(\phi) (\Lambda_{QCD}/ T)^{3.8}~~.
\label{eq:o3}
\end{equation} 
This expression must be interpolated with eq.~(\ref{eq:o2}), to study
the fluctuation onset for $T \sim \Lambda_{QCD}$. We solved the
equations of section 2 by assuming
\begin{eqnarray}
m(T,\phi) = \, m_o(\phi) (0.1^{^{1\over3.8}} \Lambda_{QCD} / T)^{3.8(1-{a~ \over a_c})}~~~
a<a_c \nonumber 
\\ 
m(T,\phi) = m_o(\phi)
~~~~~~~~~ ~~~\, ~~~~~~~~~~~~~~~~~~~~~~~~a>a_c \nonumber
\label{eq:o4} 
\end{eqnarray}
with $a_c=2.16 \cdot 10^{-12}$. With the selected value, when $T
\lesssim 0.5\, \Lambda_{QCD}$, $m(T,\phi)$ already approaches its
low--$T$ behavior $m_o(\phi)$. Let us finally outline that
eqs.~(\ref{eq:o2}) and (\ref{eq:o3}), as well as the above
interpolation, show that $m(T,\phi)\phi$ is $\phi$--independent, as is
required to obtain the equation of motion (\ref{eq:m4}). Fig.(\ref{fig:mass}) shows
the (low--)$z$ dependence of $m_o(\phi)$.
Notice the rebounce at $z \sim 10$, whose impact on halo formation
could be critical \cite{7}.

\section{The case of SUGRA potential}
When $\theta$ performs many oscillations within a Hubble time, then
$\langle \rho_{\theta,kin} \rangle \simeq \langle \rho_{\theta,pot}
\rangle$ and $\langle p_\theta \rangle \simeq 0$. By using
eqs.~(\ref{eq:m3}),(\ref{eq:m4}),(\ref{eq:kp}), it is easy to see that
\begin{equation}
\dot \rho_\theta + 3 H \rho_\theta = {\dot m \over m}\, 
 \rho_\theta
~,~ \dot \rho_\phi + 3 H (\rho_\phi+p_\phi)
= - {\dot m \over m}  \rho_\theta ~.
\label{eq:m7}
\end{equation}
When $m$ is given by Eq~(\ref{eq:o3}) , $\dot m/m = -\dot \phi/\phi -
3.8\, \dot T/T$.  At $T \simeq 0$, instead, $\dot m / m \simeq -\dot
\phi/\phi$.  Here below, the indices $_\theta$, $_\phi$ will be replaced
by $_{DM}, _{DE}$.  Eqs.~(\ref{eq:m7}) show an energy exchange between DM
and DE.
The former eq.~(\ref{eq:m7}) can then be
integrated, yielding $\rho_{DM} \propto m/a^3$. 
This law holds also when $T \ll \Lambda_{QCD}$, and then
the usual behavior $\rho_{DM} \propto a^{-3} $ is modified, becoming
\begin{equation}
\rho_{DM} a^3 \phi \simeq {\rm const}.
\label{eq:m8}
\end{equation}

Let us now assume that the potential reads
\begin{equation}
V(\phi) = (\Lambda^{\alpha+4} / \phi^\alpha) \exp (4 \pi\phi^2/m_p^2)
\label{eq:l1}
\end{equation}
(no $\theta$ dependence); in the radiative era, it will then be
\begin{equation}
\phi^{\alpha+2} = g_\alpha \Lambda^{\alpha+4} a^2 \tau^2 ~,
\label{eq:l2}
\end{equation}
with $g_\alpha = \alpha (\alpha+2)^2/4(\alpha+6)$.  This tracker
solution holds until we approach the quark--hadron transition. Then,
in Eq.~(\ref{eq:m4}), the DE--DM coupling term, $\phi \dot \theta^2$,
exceeds $a^2 V'$ and we enter a different tracking regime. 

\begin{figure}[t]
\includegraphics[height=8.0truecm,angle=0.]{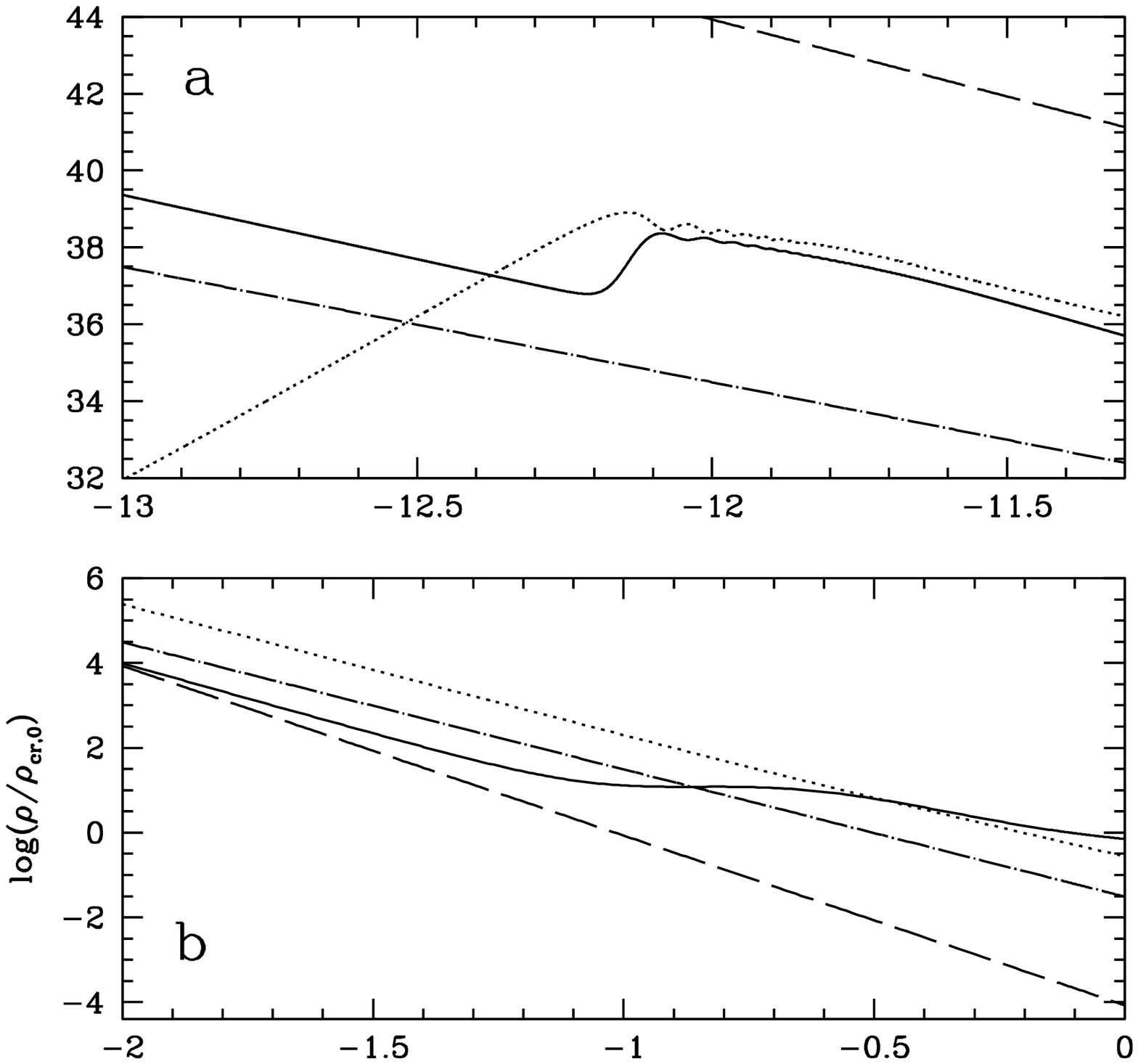}
\vskip-1.2truecm
\includegraphics[height=8.0truecm,angle=0.]{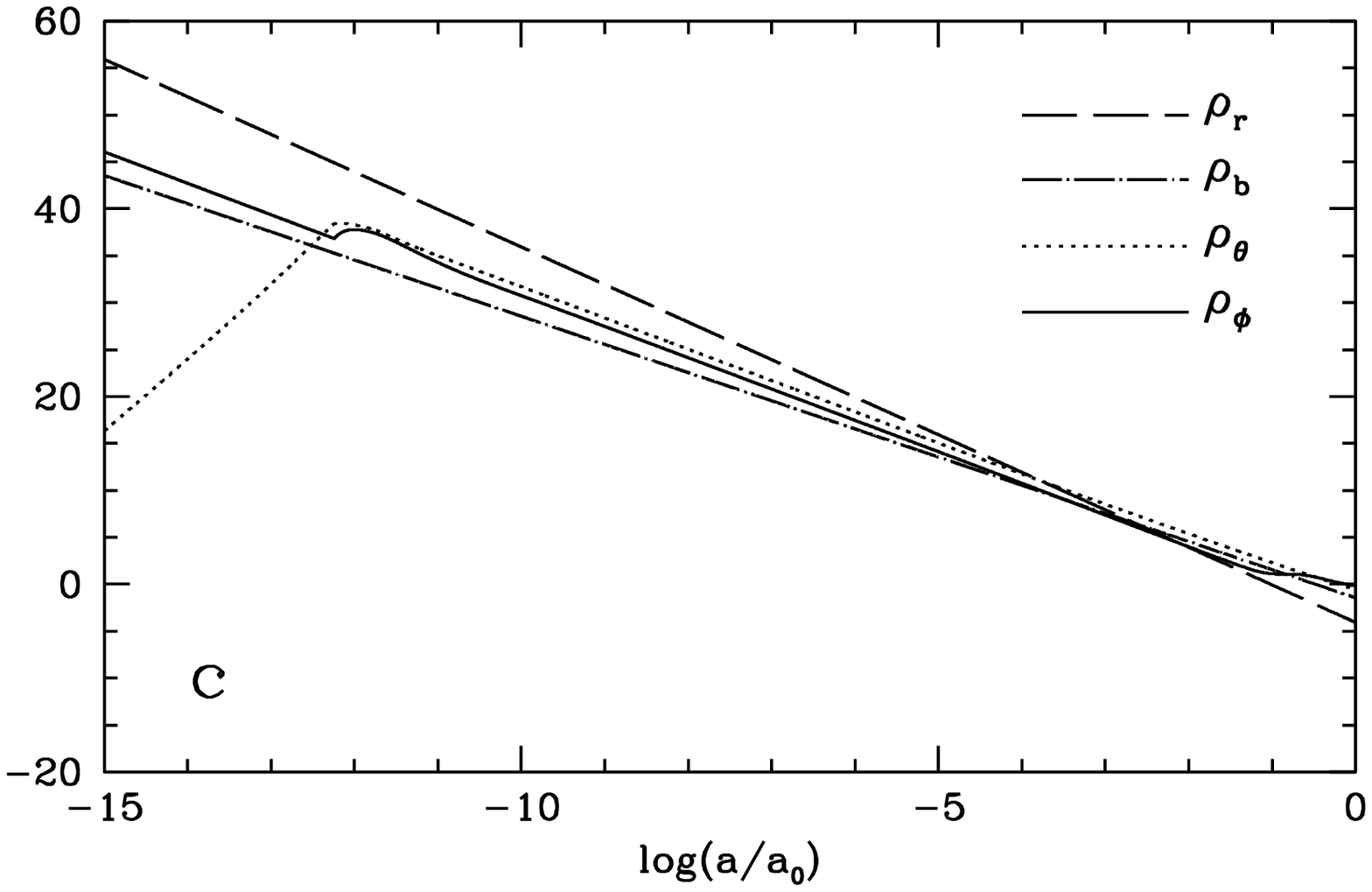}
\vskip-3.truecm
\caption{Density behaviors vs.~scale factor.}
\label{fig:rho}
\end{figure}

This is
shown in detail in Fig.(\ref{fig:rho}), 
obtained for matter (baryon) density parameters $\Omega_{m} = 0.3$
($\Omega_b = 0.03$) and Hubble constant $h = 0.7$ (in units of 100
km/s/Mpc). In particular, Fig.~(\ref{fig:rho}a) shows the transition
between these tracking regimes.  Fig.~(\ref{fig:rho}b) then shows the
low--$z$ behavior ($1+z=1/a$), since DE density exceeds radiation ($z
\simeq 100$) and then overcomes baryons ($z \simeq 10$) and DM ($z
\simeq 3$).  Fig.~(\ref{fig:rho}c) is a landscape picture for all
components, down to $a=1$.

\begin{figure}[t]
\includegraphics[height=8.truecm,angle=0.]{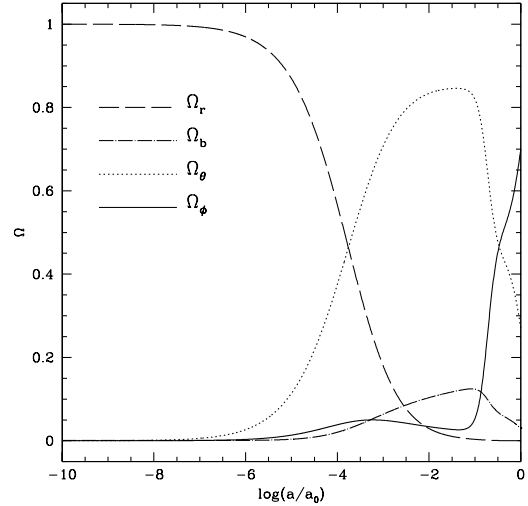}
\vskip-0.5truecm
\caption{Density parameters $\Omega_i$ vs.~scale factor $a$.}
\label{fig:omega}
\end{figure}

Notice the $a$ dependence of $\rho_{DM}$, occurring according to
Eq.~(\ref{eq:m8}). In Fig.~(\ref{fig:omega})
the related behaviors of the density parameters 
$\Omega_i$ ($i = r,~b,~\theta,~\phi$, i.e.
radiation, baryons, DM, DE) are shown.

In general, once the density parameter $\Omega_{DE}$ (at $z=0$) 
is assigned, a model with dynamical (coupled or uncoupled) DE
is not yet univocally determined. For instance, the potential
(\ref{eq:l1}) depends on  the parameters $\alpha$ and $\Lambda$ 
and one of them can still be arbitrarily fixed. Other potentials 
show similar features.

In this model such arbitrariness no longer exists.
Let us follow the behavior of $\rho_{DM}$, backwards in time,
until the approximation $\theta \ll 1$ no longer applies. 
This must approximately coincide
with the time when
\begin{equation}
2(\dot a/a + \dot \phi/\phi) \simeq m(T,\phi)\, a
\label{eq:k1}
\end{equation}
and  $\theta$ enters the oscillation regime. At that time,
according to Eq.~(\ref{eq:m8}),
which is marginally valid up to there, and taking $\theta = 1$,
\begin{equation}
\rho_{DM} \simeq \rho_{o,DM} {\phi_o \over \phi(a)} {1 \over a^3}
\simeq m^2[T(a),\phi(a)]\, \phi^2(a) ~.
\label{eq:k2}
\end{equation}
The system made by eqs.$\, $(\ref{eq:k1}) and (\ref{eq:k2}), owing to
eq.~(\ref{eq:l2}), allows to obtain (i) the scale factor $a_h$ when
fluctuations start and (ii) the scale $\Lambda$ in the potential
(\ref{eq:l1}). To do so, the present density of DM, $\rho_{o,DM}$, (or
$\Omega_{DM}$) must be assigned.  But, as we shall soon outline, the
observational value of the density in the world forces $a_h$ to
lay about the quark--hadron transition, while also $\Lambda$ is
substantially fixed.

The plots in the previous section are for
$\Omega_{DM} = 0.27$; then $\Lambda \simeq 1.5 \cdot 10^{10}$GeV and
$a_h \sim 10^{-13}$ are required by the system.  But, when
$\Omega_{DM}$ goes from 0.2 to 0.4, $\log_{10}(\Lambda/ {\rm GeV})$
(almost) linearly runs in the narrow interval 10.05--10.39$\, $, while
$a_h$ steadily lays at the eve of the quark--hadron transition.

\section{Evolution of inhomogeneities}

Besides of predicting fair ratios between the world components,
a viable model should also allow the formation of structures
in the world. 

The dual axion model belongs to the class of coupled DE models treated
by Amendola \cite{8}, with a time--dependent coupling $ C(\phi) =
1/\phi$. A $\phi$--$MDE$ phase therefore exists, after
matter--radiation equivalence, as the kinetic energy of DE is
non--neglegible during the matter--dominated era.

\begin{figure}[t]
\includegraphics[height=8.0truecm,angle=0.]{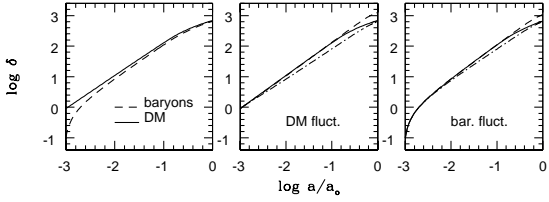}
\vskip-4.5truecm
\caption{Time evolution of DM and baryon fluctuations.}

\label{fig:flu}
\end{figure}

Fluctuation evolution is then obtained by solving the equations in
\cite{7}, with the above $C(\phi)$. The behavior shown in Fig.(\ref{fig:flu}) (left)
is then found.

Fig (\ref{fig:flu}) (center and right) compare fluctuation evolutions
the dual axion model (solid curves), with those in an analogous
$\Lambda$CDM model (dot--dashed curves) and in a coupled DE model with
constant coupling $C=0.25\, \sqrt{8\pi G} \simeq\langle C(\phi)
\rangle$ (dashed curves). As shown by the plots, the overall growth,
from recombination to now is similar in dual axion and $\Lambda$CDM
models, being quite smaller than in DE models with constant coupling.
The differences of dual axion from $\Lambda$CDM are: (i) objects form
earlier and (ii) baryon fluctuations keep below DM fluctuations until
very recently.

\section{Comparison with WMAP data}

We shall then test the dual axion model towards CMB data, together
with other dynamical or coupled DE cosmologies, by using a parameter
space of 7 to 8 dimensions. We use a Markov Chain Monte Carlo (MCMC)
approach (e.g. \cite{9}), just as in the original analysis of WMAP
first--year data \cite{10}, constraining flat $\Lambda$CDM in respect
to six parameters: $\Omega_{b} h^2$, $\Omega_{m} h^2$, $h$, $n$,
the fluctuation amplitude $A$ and the optical depth $\tau$. Notice
that, with the naming convention used here, $\Omega_{m} \equiv
\Omega_{b} + \Omega_{DM}$.

Here, three classes of DE are considered:
(i) SUGRA dynamical DE, requiring the introduction of the parameter
$\lambda = \log_{10}(\Lambda/{\rm GeV})$, the energy scale in the
potential (12). (ii) Constant coupling SUGRA DE, requiring a further
parameter, the coupling $\beta = C\,(3 m_p^2 /16 \pi)^{1/2}$. (iii)
Coupled models including the dual axion case. For this model the
parameter $\beta$ does not exist, being $C = \phi^{-1}$ and also the
scale $\Lambda$ is constrained by the requirement that $\Omega_{DE}$
lays in a fair range (also solving the strong $CP$ problem), so that
$\Lambda$ and $\Omega_{DE}$ are no longer independent
parameters. The (iii) class of model, that we call $\phi^{-1}$ models,
is however set keeping $\lambda$ as a free parameter. We aim to test
whether data constrain $\lambda$ into the right region, turning a
generic $\phi^{-1}$ model into a dual--axion model.

To use the MCMC, we need an algorithm providing the $C_l$'s. Here we
use our optimized extension of CMBFAST \cite{11}, able to inspect the
cosmologies (i), (ii) and (iii). Then, the likelihood of each model is
evaluated through the publicly available code by the WMAP team
\cite{12} and accompanying data \cite{13}.

\subsection{Results}
The basic results of our analysis are summarized in the
Table~\ref{tab:res1}, \ref{tab:res2} and \ref{tab:res3}. For each model category we list
the expectation values $\langle x \rangle$ of each parameter $x$ and the associated variance
$\sigma_x$;
we also list the values of the parameters of the best fitting models $x_{max}$.

Let us however remind that $\phi^{-1}$ models, whose fitting
results are reported in Table~\ref{tab:res3}, 
include the dual--axion model, but many other cases
as well. Our approach was meant to test whether CMB data carry
information on $\lambda$ and how this information fits the $\lambda$
range turning a $\phi^{-1}$ model into the dual--axion model.

\begin{figure}[t]
\includegraphics[height=7.0truecm,angle=0.]{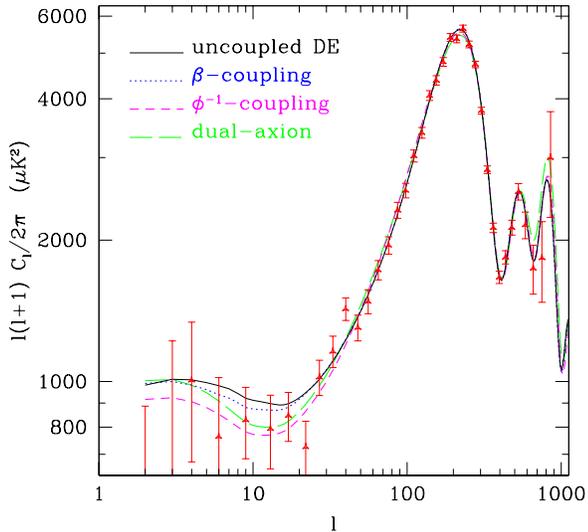}
\vskip-0.3truecm
\caption{$C_l^T$ spectra for the best fit SUGRA (solid line), constant
coupling (dotted line), $\phi^{-1}$--coupling (dashed) and dual--axion
(dot--dashed) models.} 
\label{fig:cl}
\end{figure}

At variance from uncoupled and costant-coupling SUGRA models, WMAP
data yield constraints on $\lambda$ for $\phi^{-1}$ models (see Table
\ref{tab:res3}) and the 2--$\sigma$ $\Lambda$--interval ranges from
$\sim 10$ to $\sim 3 \cdot 10^{10}$GeV, so including the dual--axion
model.
\begin{table}[t]
\caption{SUGRA parameters for uncoupled DE} 
\label{tab:res1}
\begin{center}
\begin{tabular}{|l|c|c|c|}
\hline \textbf{$x$} & \textbf{$\langle x \rangle$} & \textbf{$\sigma_x$}
 & \textbf{$x_{max}$}  
\\
\hline $\Omega_{b} h^2$  &     0.025  &   0.001  &    0.026  \\
\hline $\Omega_{DM} h^2$  &     0.12   &   0.02   &    0.11   \\
\hline $ h $           &     0.63   &   0.06   &    0.58   \\
\hline $ \tau$         &     0.21   &   0.07   &    0.28   \\
\hline $ n_s$          &     1.04   &   0.04   &    1.08   \\
\hline $ A $           &     0.97   &   0.13   &    1.11   \\
\hline $\lambda$       &     3.0    &    7.7   &    13.7   \\
\hline
\end{tabular}
\end{center}

\caption{SUGRA parameters in the presence of a constant DE--DM
coupling $\beta$.}
\label{tab:res2}
\begin{center}
\begin{tabular}{|l|c|c|c|}
\hline \textbf{$x$} & \textbf{$\langle x \rangle$} & \textbf{$\sigma_x$}
 & \textbf{$x_{max}$}  
\\
\hline $\Omega_{b} h^2$  &     0.024  &   0.001  &    0.024  \\
\hline $\Omega_{DM} h^2$  &     0.11   &   0.02   &    0.12   \\
\hline $ h $           &     0.74   &   0.11   &    0.57   \\
\hline $ \tau$         &     0.18   &   0.07   &    0.17   \\
\hline $ n_s$          &     1.03   &   0.04   &    1.02   \\
\hline $ A $           &     0.92   &   0.14   &    0.93   \\
\hline $\lambda$       &    -0.5    &    7.6   &    8.3    \\
\hline $\beta$         &     0.10   &   0.07   &    0.07   \\
\hline
\end{tabular}
\end{center}
\caption{SUGRA parameters for a $\phi^{-1}$ model.}
\label{tab:res3}
\begin{center}
\begin{tabular}{|l|c|c|c|}
\hline \textbf{$x$} & \textbf{$\langle x \rangle$} & \textbf{$\sigma_x$}
 & \textbf{$x_{max}$}  
\\
\hline $\Omega_{b} h^2$  &     0.025  &   0.001  &    0.026  \\
\hline $\Omega_{DM} h^2$  &     0.11   &   0.02   &    0.09   \\
\hline $ h $           &     0.93   &   0.05   &    0.98   \\
\hline $ \tau$         &     0.26   &   0.04   &    0.29   \\
\hline $ n_s$          &     1.23   &   0.04   &    1.23   \\
\hline $ A $           &     1.17   &   0.10   &    1.20   \\
\hline $\lambda$       &     4.8    &    2.4   &     5.7   \\
\hline
\end{tabular}
\end{center}
\end{table}
In Fig.~(\ref{fig:cl}) the $C_l^T$ spectra for all best--fit models
(apart of $\Lambda$CDM) are compared. At large $l$ all models yield
similar behaviors and this is why no model category prevails.
Discrimination could be achieved by improving large angular scale
observation, especially for polarization, so to reduce the errors on
small--$l$ harmonics.

\section{Discussion}
A first point worth outlining is that SUGRA (uncoupled) models,
bearing precise advantages in respect to $\Lambda$CDM, are consistent
with WMAP data. The ratio $w = p/\rho$ at $z=0$, for most these
models, fulfills the constraint $w \lesssim - 0.80$. However,
these models exhibit a fast $w$ variation, and $w$ becomes
$\sim$ -0.6, -0.7 at $z \sim 1$--2. In spite of this sharp decrease,
however, there is no conflict with data.

The results shown in the previous section were also considered in the
presence of some priors. For uncoupled or constant--coupling SUGRA
models, adding priors scarsely affects conclusions.
$\phi^{-1}$--models, instead, are to be discussed separately.

In the first two categories of models opacity ($\tau$) is pushed to
values even greater than in $\Lambda$CDM (see also \cite{14}).  This
can be understood in two complementary ways: (i) Dynamical DE models,
in general, exhibit a stronger ISW effect, increasing $C_l^T$ in the
low--$l$ plateau (e.g. \cite{15}). To compensate this effect, in the
fit of data, the spectral index $n_s$ is pushed to greater values. In
turn, owing to the $\tau$--$n$ degeneracy, this is compensated by
increasing $\tau$.  (ii) If the TE correlation is also considered, it
is worth reminding that dynamical DE lowers the TE correlation at low
$l$ \cite{16}. To fit the same observed correlation level, a greater
$\tau$ is therefore favored.

Greater $\tau$'s have an indirect impact also on $\Omega_b h^2$ whose
best--fit value becomes greater, although consistent with $\Lambda$CDM
within 1--$\sigma$. If we impose, as a prior, $\Omega_b h^2 = 0.0214
\pm 0.0020$ (BBNS estimates \cite{17}), $h$ shifts slightly below HST
findings, still well within 1--$\sigma$. We shall consider the effect
of a prior also on $h$.

The prior on $\Omega_b h^2$ affects reionization and the spectral
slope: $\tau$ and $n_s$ are lowered to match WMAP's findings. 
The prior on $h$, in the absence of coupling, favors low--$\lambda$
models, closer to $\Lambda$CDM. Quite in general, in fact, the sound
horizon at decoupling is unaffected by the energy scale $\Lambda$,
while the distance of the last--scattering--band is smaller for greater
$\lambda$'s. Then, as $\lambda$ increases, a lower $h$ is favored to
match the position of the first peak. In the presence of coupling,
there is a simultaneous effect on $\beta$: greater $\beta$'s yield
a smaller sound horizon at recombination, so that the distribution on
$h$ is smoother.

$\phi^{-1}$ models exhibit rather different features.  Parameters are
more strongly constrained in this case, as already outlined for the
scale $\Lambda$.  The main puzzling feature of $\phi^{-1}$ models is
that large $h$ is favored: the best--fit 2--$\sigma$ interval does not
extend below 0.85$\, $.

The problem is more severe for dual--axion $\lambda$'s values, whose
preferred values are well within 2--$\sigma$.  This model naturally
tends to displace the first $C_l^T$ peak to greater $l$ (smaller
angular scales) as coupling does in any case does. The model, however,
has no extra coupling parameter and the intensity of coupling is
controlled by the scale $\Lambda$. Increasing this scale requires a
more effective compensation and still greater values of $h$ are
favored. 

This effect appears however related to the choice of SUGRA potentials,
which is just meant to provide a concrete framework for the dual axion
model. To deepen its analysis, the contribution to the axion abundance
due to the decay of possible topological defects should also be
discussed. 

A previous analysis of WMAP limits on constant coupling models had
been carried on by \cite{19}. Their analysis
concerned potentials $V$ fulfilling the relation $dV/d\phi = BV^N$,
with suitable $B$ and $N$. They also assumed that $\tau \equiv 0.17$.
Our analysis deals with a different potential and allows more general
parameter variations. The constraints on $\beta$ we find are less
severe. It must be however outlined that $\beta > 0.1$--0.2 is
forbidden by a non--linear analysis of structure formation \cite{7}.

\section{Conclusions}
The first evidences of DM date some 70 years ago, although only in the
late Seventies limits on CMB anisotropies made evident that a
non--baryonic component had to be dominant. DE could also be dated
back to Einstein's {\it cosmological constant}, although only SNIa
data revived it, soon followed by data on CMB and deep galaxy samples.

Axions have been a good candidate for DM since the late Seventies,
although various studies, as well as the occurrence of the SN 1987a,
strongly constrained the PQ scale around values $10^{10} \lesssim
F_{PQ} \lesssim 10^{12}$GeV. Contributions to DM from topological
singularities (cosmic string and walls) narrowed the constraints to
$F_{PQ}$.  Full agreement on the relevance of such contributions has
not yet been attained and, in this paper, they are still disregarded,
while they could cause shifts in our quantitative predictions. This
point must be therefore deepened in further work.

The fact that DM and DE can both arise from scalar fields, just by
changing the power of the field in effective potentials, already
stimulated the work of various authors. A potential like (\ref{eq:l1})
was considered in the so--called {\it spintessence} model
\cite{20}. According to the choice of parameters, $\Phi$ was shown to
behave either as DM or as DE. Padmanabhan \& Choudhury \cite{21}, instead,
built a tachionic model where DM and DE arise from a single scalar
field.

Here, the scalar field $\Phi$, accounting for {\it both} DE and DM,
arises in the solution of the strong--$CP$ problem: as in the PQ
model, in eq.~(\ref{eq:n1}) $\theta$ is turned into a dynamical
variable, the phase of $\Phi$. Here, however, the modulus of $\Phi$
increases in time, approaching $m_p$ by our cosmic epoch, when it is
DE; meanwhile, $\theta$ is driven to approach zero, still performing
harmonic oscillations which are axion DM. The critical time for the
onset of coherent axion oscillations is the eve of the quark--hadron
transition, because of the rapid increase of the axion mass
$m(T,\phi)$. In our dual axion scheme, the constant $F_{PQ}$ scale of
the PQ model is replaced by the slowly varying field $\phi$.  Instead
of the scale $F_{PQ}$, data fix the scale $\Lambda$, in the SUGRA
potential. This unique setting provides DM and DE in fair proportions.
\begin{figure}[t]
\includegraphics[height=8.truecm,angle=0.]{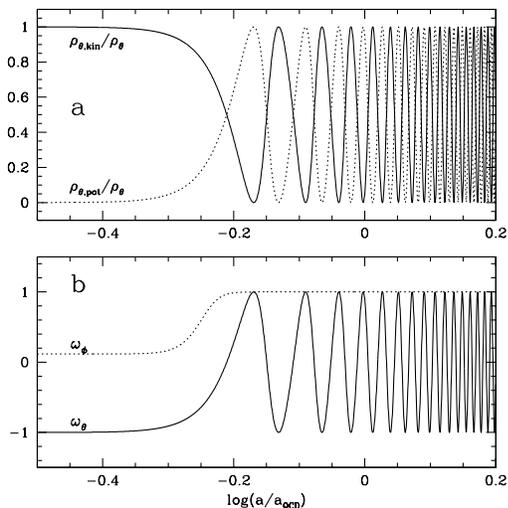}
\vskip-0.8truecm
\caption{The onset of coherent axion oscillations,
due to the increase of $m(T,\phi)$, causes the behaviors of 
$\rho_{\theta,pot}$, $\rho_{\theta,kin}$ (a) and
$\omega_{\phi,\theta} = p_{\phi,\theta}/\rho_{\phi,\theta}$ (b)
shown here.}
\label{fig:w}
\end{figure}

We therefore simultaneously solve the strong $CP$ problem and yield DM
and DE in fair proportions by setting a single parameter.



\begin{thebibliography}{}

\bibitem{1}
Peccei R.D. \& Quinn H.R. 1977, Phys.Rev.Lett. 38, 1440;
but see also: Kim J.E. 1979, Phys.Rev.Lett. 43, 103.

\bibitem{2}
Weinberg S. 1978, Phys.Rev.Lett. 40, 223; Wilczek F. 1978,
Phys.Rev.Lett. 40, 279

\bibitem{3}
Preskill J. et al 1983, Phys.Lett B120, 225; Abbott L. \& Sikivie
P. 1983  Phys.Lett B120, 133; Dine M. \& Fischler 1983 Phys.Lett B120,
137; Turner M.S. 1986 Phys.Rev.D 33, 889.


\bibitem{4}
Mainini R. \& Bonometto S.A, 2004, Phys.Rev.Lett. 93, 121301;
Mainini R., Colombo L.P.L., \& Bonometto S.A, 2005, ApJ submitted, astro-ph/0503036

\bibitem{5}
Wetterich C., 1995 A\&A 301, 32;  Ratra B. \& Peebles P.J.E., 
1988, Phys.Rev.D, 37, 3406; Ferreira G.P. \& Joyce M. 1998 Phys.Rev.D
58, 023503;
Brax, P. \& Martin, J., 1999, Phys.Lett., B468, 40; 
Brax, P. \& Martin, J., 2001, Phys.Rev. D, 61, 10350;
Brax P., Martin J., Riazuelo A., 2000, Phys.Rev. D, 62, 103505

\bibitem{6}
Kolb R.W. \& Turner M.S. 1990 The Early Universe, Addison Wesley,
and references therein

\bibitem{7}
Macci\`o A., Quercellini C., Mainini R., Amendola L., Bonometto S.A. 2004
Phys.Rev.D (in press), astro-ph/0309671; see also Mainini R., Macci\`o
A.V., Bonometto S.A., Klypin A. 2003 Ap.J. 599, 24; Klypin A., Macci\`o
A.V., Mainini. R., Bonometto S.A. 2003 Ap.J. 599, 31.

\bibitem{8}
Amendola L. 2000, Phys.Rev D 62, 043511; Amendola L. 2003 Phys.Rev.D
(subm.) and astro-ph/0311175

\bibitem{9}
Christensen N., Meyer R., Knox L. \& Luey, B. 2001, 
Classical and Quantum Gravity, 18, 2677; Knox L., Christensen, N. \& Skordis C. 2001, 
ApJ 63, L95; Lewis, A. \& Bridle, S. 2002, Phys.Rev. D 66, 103511; 
Kosowsky A., Milosavljevic M. \& Jimenez R. 2002, Phys.Rev.D 66, 63007;
Dunkley J., et al. 2004, MNRAS, submitted, astro--ph/0405462, 2004

\bibitem{10}
Spergel D.N. et al., 2003, ApJ Suppl. 148, 175

\bibitem{11}
Seljak U. \& Zaldarriaga M. 1996, ApJ, 469 437

\bibitem{12}
Verde et al. 2003, ApJ Suppl., 148, 195

\bibitem{13}
Hinshaw G., et al. 2003, ApJ Suppl., 148, 135;
Kogut A., et al. 2003, ApJ Suppl., 148, 161

\bibitem{14}
Corasaniti P.S., et al. 2004, Phys. Rev. D, in press,  
astro--ph/0406608

\bibitem{15}
Weller J. \& Lewis A.M. 2003, MNRAS, 346, 987

\bibitem{16}
Colombo L.P.L., Mainini R. \&
Bonometto S.A. 2003, in proceedings of Marseille 2003 Meeting, 'When
Cosmology and Fundamental Physics meet'

\bibitem{17}
Kirkman D. et al. 2003, ApJS, 149, 1

\bibitem{18}
Ciardi B., Ferrara A. \& White S.D.M. 2003, MNRAS, 344, L7;
Ricotti M. \& Ostriker J.P. 2004, MNRAS, 350, 359

\bibitem{19}
Amendola L. \& Quercellini C. 2003, Phys. Rev. D69, 023514

\bibitem{20}
Boyle L.A., Caldwell R.R.\& Kamionkowski M. 2002, Phys. Lett. B545, 17;
Gu Je-An \& Hwang W-Y. P. 2001, Phys. Lett. B517, 1

\bibitem{21}
Padmanabhan T., Choudhury T.R., 2002, Phys.Rev. D66, 081301.
\end{thebibliography}

{}
\end{document}